\documentstyle[12pt,fullpage,titlepage]{article}

\renewcommand{\l}{\lambda}
\renewcommand{\L}{\Lambda}
\renewcommand{\a}{\alpha}
\renewcommand{\b}{\beta}

\newcommand{\blk}{\beta_{l,k}}
\newcommand{\bllk}{\beta_{l-1,k}}
\newcommand{\blkk}{\beta_{l,k-1}}

\newcommand{\ex}{\mathop{\rm e}\nolimits}

\def\d{\delta}

\begin{document}
\rightline{hep-th/9604153}
\rightline{LANDAU-96-TMP-1}
\rightline{Appril 21, 1996}

\vfil
\centerline{\huge\bf Vertex Operators for Deformed}
\vskip 0.5cm
\centerline{\huge\bf Virasoro Algebra}
%\centerline{\huge Vertex Operators for the Deformed}
%\centerline{\huge Virasoro Algebra}
\bigskip\null\bigskip
\centerline{\large A.A.Kadeishvili}
\bigskip
\centerline{Landau Institute for Theoretical Physics,}
\centerline{Kosygina 2, GSP-1, 117940 Moscow V-334, Russia}
\centerline{\sl e-mail: kadeishv@itp.ac.ru}
\bigskip\null\bigskip

\centerline{\large\bf Abstract}
\bigskip
Vertex operators for the deformed Virasoro algebra are defined,
their bosonic representation is constructed
and difference equation for the simplest vertex operators
is described.

\vfil\vfil\eject

\section{Introduction}
A feature characterising both classical and quantum integrable
systems is the existence of the infinite dimensional Abelian
symmetry. For two-dimensional lattice systems this symmetry appears
as a family of commuting transfer matrices. From the standpoint of
one-dimensional quantum chains this means that there are infinitely
many mutually commuting integrals of motion.

There is a well known method for studying the eigenvalues of commuting
transfer matrices. The Bethe Ansatz method makes it possible to obtain
all eigenvalues and eigenvectors in a certain  form ,
using solution to a system of algebraic equations (the Bethe Ansatz
equations). With this approach  the main problem is to handle the
Bethe Ansatz equations in the thermodynamic limit when the number of sites
of the lattice tends to infinity.

Fortunately, in this limit solvable lattice models have
another type of symmetry usually called dynamical one \cite{book},
\cite{LukPug}.
For six-vertex model this symmetry is described by quantum affine
group $U_q(\widehat{sl}_2)$ at the level one \cite{book}.
Similar picture appears when considering ABF model. In this case
dynamical symmetry is described by Deformed Virasoro Algebra (DVA)
\cite{LukPug}.

Generators of the dynamical symmetry do not mutually commute;
however they form the space of states of the model. The role of this
symmetry in the lattice models is the same as that of the Virasoro
algebra in  Conformal Field Theory (CFT). Thus methods and results
of CFT \cite{BPZ} may be used in attacking lattice models.

In CFT space of physical states is defined by irreducible representations
of Virasoro algebra. Vertex operators intertwine this representations.
Physicaly, operators correspond to primary fields of CFT. The overall
space of states of the theory can by obtained
by action of the vertex operators on the vacuum state. The bosonization is
a powerful method of CFT \cite{FeFu,DotsFat,Feld}.
It gives an explicit description of the space of
states of the theory and enables one to calculate correlation
functions of all physical fields of the theory.

In \cite{LukPug2,Jap} a bosonic representation for the DVA was described.
A bosonic representation for the simplest vertex operatos was given in
\cite{LukPug2}. Using this results, correlation functions of the ABF
model were calculated in \cite{LukPug}.

In this study we give an abstract definition of vertex operators for the DVA
and describe their bosonic representation. Our construction is in line with
that of \cite{Feld}. In conclusion we suggest an application
of our construction to the fusion RSOS models \cite{Jimb} and discuss
problems to be solved.

\section{Deformed Virasoro Algebra}

Let us recall some basic facts on the DVA \cite{LukPug2,Jap,FeFr}.
The algebra depends on the two parameters $\xi>0$ and $0<x<1$.
The algebra is generated by the current $T(z)=\sum T_n z^{-n}$ with the
following basic relations
\begin{eqnarray}
&&f\left({w\over z}\right)T(z)T(w)-f\left({z\over w}\right)T(w)T(z)\cr
&&\quad={\left(x^\xi-x^{-\xi}\right)\left(x^{(\xi+1)}-x^{-(\xi+1)}\right)
\over x+x^{-1}}\left(\delta\left({w \over z}x^{-2}\right)-
\delta\left({w \over z}x^2\right)\right),\label{1}
\end{eqnarray}
where
$$
f(z)=(1-z)^{-1}{\left(zx^{2(\xi+1)};x^4\right)_\infty
			   \left(zx^{-2\xi};x^4\right)_\infty
		  \over\left(zx^{2(\xi+1)+2};x^4\right)_\infty
			   \left(zx^{-2\xi+2};x^4\right)_\infty},
$$
$$
(z;q_1,...q_p)_\infty=\prod_{n=0}^\infty (1-z q_1^n)...(1-z q_p^n)
$$
and
$$
\delta(z)=\sum_{m=-\infty}^{\infty}z^m
$$
If we fix z and consider the limit as $x\to 1$, then
\begin{equation}
T(z)=2+\xi(\xi+1)(x-x^{-1})^2\left(z^2 L(z)+
{1\over 4\xi(\xi+1)}\right)+O\left((x-x^{-1})\right).
\end{equation}
It can be verified that the defining relations (\ref{1}) give Virasoro
algebra commutators for the current $L(z)$ and that the corresponding central
charge is
\begin{equation}
c=1-{6\over \xi(\xi+1)}=1-12\a_0^2,
\end{equation}
where we introduce the notation similar to those used in CFT
\begin{eqnarray}
\a_0&={\textstyle 1\over \textstyle \sqrt{2\xi(\xi+1)}},\nonumber\\
\a_+&=-\sqrt{\textstyle 2(\xi +1)\over \textstyle \xi},\\
\a_-&=\sqrt{\textstyle 2\xi\over \textstyle \xi+1}.\nonumber
\end{eqnarray}

The DVA may be bosonized in terms of deformed bosonic oscillators
$\l_m$ with the commutation relations
\begin{equation}
[\l_m,\l_n]={1\over m}{\left(x^{\xi m}-x^{-\xi m}\right)
   \left(x^{(\xi+1)m}-x^{-(\xi+1)m}\right)
 \over x^m + x^{-m}}\delta_{n+m,0}.\label{2}
\end{equation}

We should also define "zero mode" operators $P,Q$ which commute
with $\l_m$ and satisfy the relation:
\begin{equation}
\left[Q,P\right]=i.\label{3}
\end{equation}
The Heisenberg  algebra (\ref{2}), (\ref{3}) is represented on the Fock
space ${\cal F}_p$ generated by the action of the creation operators
$\l_{-n}$, $n>0$, on the highest weight vector $\upsilon_p$
\begin{equation}
\l_n \upsilon_p=0,\quad n\ge 0, \qquad P\upsilon_p=p\upsilon_p.
\end{equation}

Now we introduce the field
\begin{equation}
\L(z)=x^{\sqrt{2\xi(\xi+1)}P}:\exp\biggl(-\sum_{m\ne 0}\l_m z^{-m}\biggr):.
\end{equation}
Then the current T(z) may be written in the form \cite{Jap}, \cite{FrResh}
\begin{equation}
T(z)=x\L(zx^{-1})+x^{-1}\L^{-1}(zx).\label{4}
\end{equation}
It can be easily verified that the field $T(z)$ satisfies (\ref{1}).

In what follows of interest is a particular representations
of DVA which correspond to minimal models of CFT \cite{BPZ}. This
representations are parametrized by two positive coprime integer numbers
$p$ and $p\prime$ $(p\prime > p)$. In this parametrization we have
\begin{equation}
\xi={p\over p\prime-p},\quad \a_+=-\sqrt{2{p\prime\over p}},
\quad \a_-=\sqrt{2{p\over p\prime}}.\label{5}
\end{equation}
Let us consider the action of the DVA on the Fock spaces
\begin{equation}
{\cal F}_{l,k}={\cal F}_{\a_{l,k}},
\end{equation}
where $\a_{l,k}=-{1\over 2}(\a_+l+\a_-k)$, $k,l\in {\bf Z}$.

This action is highly reducible and some states should be excluded
from the Fock module to obtain irreducible component. Explicitly,
the procedure is as follows \cite{LukPug},
\cite{Feld}.
Let us introduce "screening current"
\begin{equation}
I_+(z)=\ex^{i\a_+Q}z^{\a_+P+1}
:\exp\left\{-\sum_{m\ne 0}
{x^m+x^{-m}\over x^{\xi m}-x^{-\xi m}}\l_m z^{-m}\right\}:
\end{equation}
and the auxiliary function
$$
F_+(z,P)=z^{-\a_+P-{\xi+1\over \xi}}x^{\xi\a_+P(\a_+P+1)+\a_+P}
{E(x^{-1-2\xi\a_+P}z;x^{2\xi})\over E(x^{-1}z^{-1};x^{2\xi})},
$$
where
$$
E(z;q)=(q;q)_\infty (z;q)_\infty (z^{-1}q;q)_\infty.
$$
The "screening charge"
\begin{equation}
Q_+=\oint dz I_+(z) F_+({z\over z_0};P).\label{6}
\end{equation}
is of crucial importance in our analysis.

An important property of operator $Q_+$ is that it commutes with the
DVA current $T(z)$ and does not depend on $z_0$ acting on the Fock spase 
${\cal F}_{l,k}$ .
We note that, in view of parametrization (\ref{5}),
$\a_{l,k}=\a_{l+p,k+p\prime}$; hence the Fock modules ${\cal F}_{l,k}$ and
${\cal F}_{l+p,k+p\prime}$ may be identified.

The screening charge (\ref{6}) determines the following map:
\begin{eqnarray}
Q_{2j}&=Q_+^l:\quad &{\cal F}_{l-2jp,k}\to {\cal F}_{-l-2jp,k},\\
Q_{2j+1}&=Q_+^{p-l}:\quad &{\cal F}_{-l-2jp,k}\to {\cal F}_{l-2(j+1)p,k}
\end{eqnarray}
for $j\in {\bf Z}$, $1\leq l\leq p-1$, $1\leq k \leq p\prime-1$.

As a result , we can construct the infinite chain
$$
...\stackrel{Q_{-2}}{\longrightarrow}{\cal F}_{2p-l,k}
\stackrel{ Q_{-1}}{\longrightarrow}{\cal F}_{l,k}
\stackrel{Q_0}{\longrightarrow}{\cal F}_{-l,k}\stackrel{Q_1}...
$$

It was shown in \cite{LukPug} that this chain is the Felder resolution
\cite{Feld}. In other words
$$
\begin{array}{l}
1. Q_j Q_{j-1}=0\\
2. Ker Q_j/Im Q_{j-1}=\left\{\begin{array}{ll}
			      {\cal L}_{l,k}, \quad &j=0,\\
			      0,& j\neq0,
			      \end{array}\right.
\end{array}
$$
where ${\cal L}_{l,k}$ is an irreducible representation of the DVA.

\section{Vertex Operators for the DVA}

In the previous section we described the irreducible representations of the 
DVA. Now we construct vertex operators intertwining this representations.
Our construction is parallel to that of \cite{Feld}.
At first we introduce second "screening charge"
\begin{equation}
Q_-=\oint dz I_-(z) F_-({z\over z_0};P),\label{7}
\end{equation}
where
\begin{equation}
I_-(z)=\ex^{i\a_-Q}z^{\a_-P+1}
:\exp\left\{\sum_{m\ne 0}
{x^m+x^{-m}\over x^{(\xi+1)m}-x^{-(\xi+1)m}}\l_m z^{-m}\right\}:
\end{equation}
and
\begin{equation}
F_-(z,P)=z^{-\a_-P-{\xi\over \xi+1}}x^{(\xi+1)\a_-P(\a_-P+1)-\a_-P}
{E(x^{1-2(\xi+1)\a_-P}z;x^{2\xi+2})\over E(x z^{-1};x^{2\xi+2})}.
\end{equation}
The operator $Q_+$ commutes with the DVA current $T(z)$ and the charge $Q_-$.

Then we introduce highest component of the vertex operator
\begin{equation}
V_{l,k}=\ex^{i\a_{l,k} Q}z^{\a_{l,k} P}:\exp\left\{\sum_{m\ne 0}
{x^{\blk m}-x^{-\blk m}\over
\left(x^{\xi m}-x^{-\xi m}\right)
\left(x^{(\xi+1)m}-x^{-(\xi+1)m}\right)}\l_m z^{-m}\right\}:,\label{8}
\end{equation}
where $\blk=\sqrt{2\xi(\xi+1)}\a_{l,k}=(\xi+1)l-\xi k$,
$1\leq l\leq p-1$, $1\leq k\leq p\prime-1$ .

The other components of the vertex operator are given by the action
of the screening charges on the highest component
\begin{eqnarray}
V_{l,k}^{m,n}(z)&=&\int V_{l,k}(z)
\prod_{i=1}^m dw_i I_+(w_i) F_+({w_i\over z}x^\bllk;P)
\prod_{j=1}^n du_j I_-(u_j) F_-({u_j\over z}x^{-\blkk};P)\nonumber\\
&=&\int \prod_{i=1}^m dw_i F_+({z\over w_i}x^\bllk;-P) I_+(w_i)
\prod_{j=1}^n du_j  F_-({z\over u_j}x^{-\blkk};-P) I_-(u_j)V_{l,k}(z),
\label{9}
\end{eqnarray}
where $0\leq m\leq l$, $0\leq n\leq k$ and contours for $w_i$, $u_j$
enclose the poles $w_i=z x^{-\b_{l,k+1}+2\xi q}$,
$u_j=z x^{\b_{l+1,k}+(2\xi+2)q}$, $(q=0,1,2...)$ respectevly .

The vertex operator $V_{l,k}^{m,n}(z)$ maps Fock space ${\cal F}_{s,t}$
into ${\cal F}_{s+l-2m,t+k-2n}$.

Note that operators $V_{1,0}$ and $V_{0,1}$ coincide accurate to normalization
factors with the vertex operators constructed in \cite{LukPug}, \cite{LukPug2}.
In some sense these operators are fundamental, because any other vertex
operator may be obtained as their product
\begin{equation}
V_{l,k}^{m,n}(z)=
:\prod_{i=-l/ 2}^{l/2}V_{1,0}^{m_i,0}\left(x^{2i(\xi+1)}z\right)
\prod_{j=-k/2}^{k/2}V_{0,1}^{0,n_j}\left(x^{2j\xi}z\right):\label{10}
\end{equation}
here $m=\sum m_i$, $n=\sum n_j$.

To prove the intertwining property of the vertex operators we should test
the following diagram
$$
\begin{array}{ccc}
{\cal F}_{s,t}&\stackrel{V_{l,k}^{m,n}(z)}{\longrightarrow}&
{\cal F}_{s+l-2m,t+k-2n}\\
\downarrow\lefteqn{Q_+^t}&&\downarrow\lefteqn{Q_+^{t+k-2n}}\\
{\cal F}_{s,-t}&\stackrel{V_{l,k}^{m,k-n}(z)}{\longrightarrow}&
{\cal F}_{s+l-2m,2n-t-k} \end{array}
$$
for commutativity.
This commutativity is a consequence of the
following basic properties of the screening charges
\begin{eqnarray}
Q_+^t V_{l,k}^{m,n}(z)&=(-1)^{t-l}V_{l,k}^{m,n+t}(z)\nonumber\\
V_{l,k}^{m,n}(z)Q_+^t&=V_{l,k}^{m,n+t}(z)\nonumber.
\end{eqnarray}
Thus operator (\ref{9}) intertwines the irreducible representations
of the DVA.

Using bosonic representation for $T(z)$ and $V_{l,k}^{m,n}(w)$
we may calculate commutation relations between them. We restrict
our attention to the highest component $V_{l,k}(w)$ of vertex operator in
view of the commutativity of $T(z)$ with the screening charges. Thus we obtain
\begin{eqnarray}
&&{w\over z}f_{l,k}\left({w\over z}\right)T(z)V_{l,k}(w)+
f_{l,k}\left({z\over w}\right)V_{l,k}(w)T(z)\nonumber\\
&&\qquad=g_{l,k}\biggl\{
\left(wx^{-\xi-1}\right)^{-{\b_{l-1,k}\over 2\xi}}
x\d\left({w\over z}x^{\blk}\right)
V_{l-1,k}\left(wx^{-\xi-1}\right)V_{1,0}\left(wx^{\b_{l-1,k-2}}\right)\nonumber\\
&&\qquad+\left(wx^{-\b_{l-1,k-2}}\right)^{-{\b_{l-1,k}\over 2\xi}}
x^{-1}\d\left({w\over z}x^{-\blk}\right)
V_{1,0}\left(wx^{-\b_{l-1,k-2}}\right)V_{l-1,k}\left(wx^{\xi+1}\right)
\biggr\}\nonumber\\
&&\qquad=g_{l,k}^\prime\biggl\{
\left(wx^\xi\right)^{\b_{l,k-1}\over 2\xi+2}
x\d\left({w\over z}x^{\blk}\right)V_{l,k-1}\left(wx^\xi\right)
V_{0,1}\left(wx^{-\b_{l-2,k-1}}\right)\nonumber\\
&&\qquad+\left(wx^{-\b_{l-2,k-1}}\right)^{\b_{l,k-1}\over 2\xi+2}
x^{-1}\d\left({w\over z}x^{-\blk}\right)
V_{0,1}\left(wx^{-\b_{l-2,k-1}}\right)V_{l,k-1}\left(wx^{-\xi}\right)
\biggr\}, \label{11}
\end{eqnarray}
where
$$
f_{l,k}(z)={\left(zx^{\blk+2};x^4\right)_\infty
         \left(zx^{-\blk+4};x^4\right)_\infty
    \over\left(zx^{\blk};x^4\right)_\infty
         \left(zx^{-\blk+2};x^4\right)_\infty},
$$

$$
g_{l,k}={\left(x^{4\xi+4};x^{2\xi},x^4\right)_\infty
			\left(x^{2\b_{l,k-1}};x^{2\xi},x^4\right)_\infty
	   \over\left(x^{4\xi+2};x^{2\xi},x^4\right)_\infty
			\left(x^{2\b_{l+1,k}};x^{2\xi},x^4\right)_\infty}
$$
and
$$
g_{l,k}^\prime={\left(x^{-4\xi};x^{-2(\xi+1)},x^4\right)_\infty
			\left(x^{2\b_{l-1,k}};x^{-2(\xi+1)},x^4\right)_\infty
	   \over\left(x^{-4\xi-2};x^{-2(\xi+1)},x^4\right)_\infty
			\left(x^{2\b_{l,k+1}};x^{-2(\xi+1)},x^4\right)_\infty}.
$$

We mayn define a vertex operator as the intertwining operator for the
representations of the DVA with defining relations (\ref{11}).

It can be verified that, as $x\to 0$ our construction gives vertex
operators for the Virasoro algebra. In particular, operator (\ref{9}) gives
the bosonic representation for the primary field with the conformal dimension
\begin{equation}
\Delta_{l,k}={((l-1)p-(k-1)p\prime)^2-(p-p\prime)^2 \over 4pp\prime}
\end{equation}
and equation (\ref{11}) is rearranged to give
\begin{equation}
[L(z),V_{l,k}^{n,m}(w)]=z^{-1}\delta\left(w\over z\right)
{\partial\over\partial w}V_{l,k}^{n,m}(w)+
z^{-2}\Delta_{l,k}\delta\prime\left(w\over z\right)V_{l,k}^{n,m}(w).
\end{equation}
which coincides with defining relations for a primary field in CFT.

\section{Conclusion}

In the previous sections we defined vertex operators for the DVA and 
constructed their bosonic representations. Now let us review some 
our further results and problems.

1.In CFT Fock module ${\cal F}_{1,0}$ contains a cosingular vector
at the second level. This results in the differential equation of the 
second order for the vertex operator $V_{1,0}(z)$
\begin{equation}
\partial^2 V_{1,0}(w)={3\over 4\Delta_{1,0} +2}\oint_w
{L(z)V_{1,0}(w)\over z-w}dz.\label{eq1}
\end{equation}
Using this equation and the Ward identities we can derive differential
equations for correlation functions in CFT.

A similar situation emerges in the DVA. In this case the vertex operator 
$V_{1,0}(z)$ satisfies difference equation of the second order
\begin{equation}
{\cal D}_{x^{2\xi}}^2V_{1,0}(w)={1\over (x^\xi-x^{-\xi})^2}\left(
\oint_w z^{-3}f_{1,0}\left({w\over z}\right)T(z)V_{1,0}(w)dz-
(x^\xi-x^{-\xi})w^{-2}V_{1,0}(w)\right).\label{eq}
\end{equation}
Up to now, we do not know the Ward identities for the DVA, and it is
unclear how to to obtain a difference equation for correlation functions
of the DVA.

2. In \cite{LukPug} the vertex operator
$V_{0,1}(z)$ was interpreted as a semi-infinite transfer matrix
for the ABF model (in this case we should put $p\prime=p+1$). Using
the coner transfer matrix method and bosonic representationt for the
vertex operator correlation, functions of the ABF model in the regime III
were calculated. From representation (\ref{10}) and
exchange relations for operators $V_{0,1}(z)$ \cite{LukPug} it follows
that operators $V_{0,n}(z)$ give semi-infinite transfer matrices for
the fusion RSOS models described in \cite{Jimb}. The bosonic representation
(\ref{9})
makes it possible to calculate correlation functions of the fusion RSOS
models following the lines of \cite{LukPug}.
We will concentrate on this problem in a futher publications.

\bigskip
\centerline {\large\bf Acknowledgements}
\bigskip
I would like to thank A.Belavin, B.Feigin, M.Lashkevich, S.~Parkhomenko,
V.Postnikov, Ya. Pugai for discussions and encouragement.

This work supported in part by ISF grant M6N000 and RBRF grant 93-02-3135.

\end{document}